\documentclass[journal]{IEEEtran}

\usepackage{cite}     

\usepackage{graphicx}

\usepackage{amsmath}   

\begin{document}

\title{Engineering Exchange Coupling in Double Elliptic Quantum Dots}

\author{\authorblockN{L.-X. Zhang, D. V. Melnikov, and J.-P. Leburton}
\authorblockA{Beckman Institute for Advance Science and Technology\\
and Department of Electrical and Computer Engineering\\
University of Illinois at Urbana-Champaign}
}

\author{L.-X.~Zhang,
        D.~V.~Melnikov,
        and~J.-P.~Leburton
\thanks{L.-X.~Zhang is with 
Beckman Institute for Advance Science and Technology
and Department of Electrical and Computer Engineering, 
University of Illinois at Urbana-Champaign, Urbana, IL 61801 USA
(phone: 217-244-1964; fax: 217-244-4333; email: lzhang7@uiuc.edu)}
\thanks{D.~V.~Melnikov is with 
Beckman Institute for Advance Science and Technology 
and Department of Electrical and Computer Engineering,
University of Illinois at Urbana-Champaign, Urbana, IL 61801 USA
(phone:217-244-6913; fax: 217-244-4333; email: dmm3@uiuc.edu)}
\thanks{J.-P.~Leburton is with 
Beckman Institute for Advance Science and Technology
and Department of Electrical and Computer Engineering, 
University of Illinois at Urbana-Champaign, Urbana, IL 61801 USA
(phone: 217-333-6813; fax: 217-244-4333; email: leburton@ceg.uiuc.edu)}}

\maketitle

\begin{abstract}
Coupled elliptic quantum dots with different aspect ratios containing up to two electrons are studied 
using a model confinement potential in the presence of magnetic fields. Single and two particle 
Schr\"{o}dinger equations are solved using numerical exact diagonolization
to obtain the exchange energy and chemical potentials. As the ratio
between the confinement strengths in directions perpendicular and parallel to the coupling direction of the double dots increases, the exchange energy at zero magnetic field 
increases, while the magnetic field of the singlet-triplet transition
decreases. By investigating the charge stability diagram, we find that as inter-dot
detuning increases, the absolute value of the exchange energy increases
superlinearly followed by saturation. This behavior is attributed to
the electron density differences between the singlet and triplet states in the assymetric quantum dot systems. 
\end{abstract}


\section{Introduction}

Coupled quantum dots (QDs) are of particular interest for spin-based quantum computation because quantum logic gates (such as a Controlled-NOT gate) can be realized via the interaction between two spin quantum bits (qubits), i.e., the spins of two electrons, each trapped in an individual quantum dot\cite{Loss}. In such devices, the interaction between the two spins is proportional to the exchange energy $J$, which is equivalent to the splitting between the lowest singlet and triplet two-electron states. 

Extensive theoretical and experimental works have been done to study the exchange interaction in coupled semiconductor QD systems. From a theoretical point of view, variational methods such as the Heitler-London method or exact diagonalization are commonly used to obtain the dependence of the exchange interaction on the system parameters, e.g., the inter-dot separation, the tunneling barrier between the QDs, and the externally applied magnetic field \cite{Burkard,Hu,Harju,Szafran}. Experimentally, the extraction of the exchange interaction relies on the charge stability diagram in which the boundaries between distinctive stable charge states, 
i.e., between the states with fixed number of electrons ($N_1$, $N_2$) in each of the coupled dots, are represented as functions of the two controlling gate biases, one for each dot \cite{Wiel}. To obtain $J$, two different methods were used: one is based on the Hubbard model, which involves the analysis of double-triplet point (DTP) separation on the stability diagram \cite{Hatano}; the other utilizes Zeeman splitting to measure the exchange energy as a function of inter-dot detuning in a coherent control cycle of state preparation, spin-interaction, and projective readout in laterally coupled semiconductor quantum dot systems \cite{Petta}. In such an experiment, the charge state of the two qubits is manipulated through a charging cycle from the $(0,2)$ state to the $(1,1)$ state and finally back to the $(0,2)$ state in a coherent way. These kinds of experiments are stimulating interest in the investigation of the exchange interaction and the associated charge re-distribution as a function of inter-dot bias detuning in coupled QDs. 

In experimental planar laterally coupled quantum dot device structures, the strength of the  confining potentials in the coupling direction and the direction perpendicular to it in the plane of the two-dimensional electron gas are different due to the top gate patterning \cite{Zhang0, Dmm0}. 
In this paper, we account for this effect and study coupled quantum dots with different aspect ratios in the confinement potential, i.e., with different ratios of the confinement strengths in the two directions.

We introduce our model Gaussian-shaped confinement potential and exact diagonalization method in Section II. In Section III, we discuss the different dependences of the exchange interaction on the magnetic fields for different dot deformations. We then analyze the stability diagram for various coupled QD systems followed by a comparison of the dependences of the exchange energy on inter-dot detuning for dots with different aspect ratios and their relationship to electron density localization. Finally, we present concluding remarks in Section IV.

\section{Model and Method}

We use the following model potential to describe the coupled quantum dot system \cite{Szafran}:
\begin{equation}
V({\bf r}) = -V_Le^{-(x+d/2)^2/R_x^2+y^2/R_y^2}
             -V_Re^{-(x-d/2)^2/R_x^2+y^2/R_y^2},
\label{eqn:potential}
\end{equation}
where $V_L$ and $V_R$ are the depth of the left and right dots (equivalent to the QD gate voltages in experimental structures \cite{Wiel}) which can be independently varied, $d$ is the inter-dot separation, $R_x$ and $R_y$ are the extension of the each dot in the $x$ and $y$ direction, respectively.

Our computational approach consists of two steps \cite{Dmm}. In the first step, 
we solve a single-particle problem with the Hamiltonian
\begin{equation}
h({\bf r}) = \frac{1}{2m^*}({\bf p}+\frac{e}{c}{\bf A})^2 + V({\bf r}).
\label{eqn:single_H}
\end{equation}
Here, $m^* = 0.067m_e$ is the electron effective mass in GaAs, and
${\bf A} = \frac{1}{2}[-By,Bx]$ is the vector potential of the magnetic field $B$ oriented perpendicular to the $xy$-plane. This Hamiltonian
is diagonalized in the basis of the product of two harmonic oscillator
states in each direction, $\psi_i({\bf r}) = \phi_n(x)\phi_m(y)$, where
$\phi_n$ ($\phi_m$) denotes the $n$-th ($m$-th) harmonic oscillator states (in this work, we use
eight harmonic states in each direction). 

In the second step, we solve the two-particle problem for which the Hamiltonian 
is given by
\begin{equation}
H({\bf r_1},{\bf r_2}) = h({\bf r_1})+h({\bf r_2})+C({\bf r_1},{\bf r_2}),
\label{eqn:double_H}
\end{equation}
where $C({\bf r_1},{\bf r_2}) = \displaystyle{\frac{e^2}{\epsilon}\frac{1}{|{\bf r_1}-{\bf r_2}|}}$
accounts for  
the Coulomb interaction between the two electrons and $\epsilon=12.9$
is GaAs dieletric constant. The Zeeman effect is not included in (3) since its effect trivially lowers the energy of the triplet state by $\sim25$ meV/T while leaving the 
energy of the singlet state unaltered so that it can be readily accounted for later, if necessary.

The diagonalization procedure for the two-particle Hamiltonian is performed by expanding two-electron spinless wave function in the basis
\begin{equation} 
\Psi_S({\bf r_1},{\bf r_2}) = \displaystyle\sum_{ij}
\beta_{ij}[\psi_i({\bf r_1})\psi_j^*({\bf r_2})+(-1)^S\psi_j({\bf r_1})\psi_i^*({\bf r_2})],
\label{eqn:twowf}
\end{equation}
which is symmetric for singlet ($S=0$) state and antisymmetric for triplet ($S=1$) state. The summation is carried over $i \leq j$ for the singlet and $i<j$ for the triplet. 

From the two-particle wave function, we compute the electron density as

\begin{equation} 
\rho_s({\bf r_1}) = \int|\Psi_S({\bf r_1},{\bf r_2})|^2d{\bf r_2}. 
\label{eqn:density}
\end{equation}

The chemical potential of the structure is related to the total energy 
of the system as~\cite{Wiel}
\begin{equation}
\mu(N_1+N_2) = E_G(N_1+N_2) - E_G(N_1+N_2-1),
\label{eqn:mu}
\end{equation}
where $E_G(N)$ [note that $E_G(0)=0$] is the ground state energy of the $N$-electron state.
The exchange energy is given by
\begin{equation}
J = E_G^T(2) - E_G^S(2),
\label{eqn:J}
\end{equation}
where $E_G^T(2)$ and $E_G^S(2)$ denote the ground state energy for the triplet and
singlet state, respectively.

\section{Results}

\begin{figure}[t]
\centering
\includegraphics[width=2.8in]{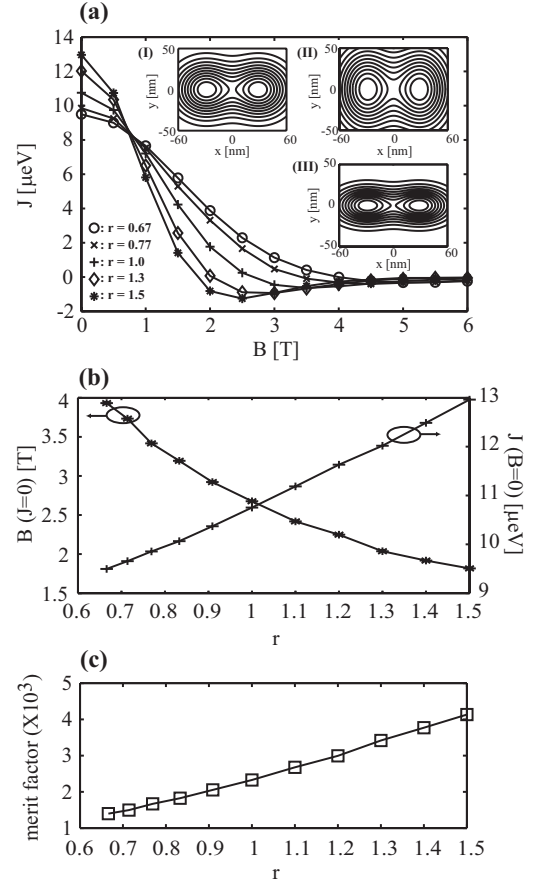}
\caption{(a) Main panel: exchange energy dependence on the magnetic field for dots with different aspect ratios $r=R_y/R_x$,  $R_x=30$ nm. The insets (I), (II) and (III) show the contour plots of the confining potential for $r=1$, $r=1.5$ and $r=0.67$, respectively. (b) Exchange energy maximum [J(B=0)] and magnetic field at singlet-triplet transition [B(J=0)] and (c) merit factor $J/\hbar\omega_c$ ($\omega_c=eB/m_*c$) as a function of the aspect ratio $r=R_y/R_x$. For (a), (b) and (c), $V_L=V_R = 25$ meV, $d=60$ nm.}
\label{fig:fig1}
\end{figure}

In Fig. \ref{fig:fig1}(a), main panel, we plot the exchange energy as a function of the magnetic
field for dots with different aspect ratio $r=R_y/R_x$, $R_x=30$ nm. In insets (I), (II), and (III),
the contour plots of the two-dimensional potential are displayed for
$r = 1$, $r = 1.5$, and $r = 0.67$, respectively. Other parameters
for the model potential are $V_L=V_R = 25$ meV, $d=60$ nm. 
It is observed that the magnitudes of maximum and minimum values of the exchange 
energy are larger for a larger aspect ratio $r$ because of enhanced inter-dot coupling. 
As the aspect ratio $r$ increases, the magnetic fields at the singlet-triplet transition point,
at the exchange energy minimum, and at the saturation point of exchange energy all shift to smaller values, leading to a more compressed appearance of the curve in the
horizontal direction. In Fig. \ref{fig:fig1}(b), as the aspect ratio increases from $0.67$ to $1.5$, the exchange energy 
maximum [$J(B=0)$] ramps up slowly from $~9.5$ to $13$ $\mu$eV, while the magnetic field at 
the singlet-triplet transition point [$B(J=0)$] decreases more rapidly from $~4$ to $1.75$ T.
For spin-based quantum computing, a larger $J(B=0)$ and a smaller $B(J=0)$ are both
desirable because they offer better control of the quantum logic gate
\cite{Loss, Petta}.

Therefore, we define a merit factor, $J(B=0)/\hbar\omega_c$, 
where $\omega_c=eB/m_*c$ is the cyclotron frequency. In Fig. \ref{fig:fig1}(c), 
we observe that the merit factor increases almost linearly with the aspect ratio, with
an enhancement factor of more than twice, as the aspect ratio increases from $0.67$ to $1.5$.  
It is also interesting to note in Fig. \ref{fig:fig1}(a) that at $B \approx 0.8$ T, the values of the exchange energy for dots of 
different aspect ratios are very close to each other ($J \approx 8.5$ $\mu$eV), which implies that at this magnetic field
geometric factor plays a minor role in determining the exchange energy.

\begin{figure}[t]
\centering
\includegraphics[width=2.7in]{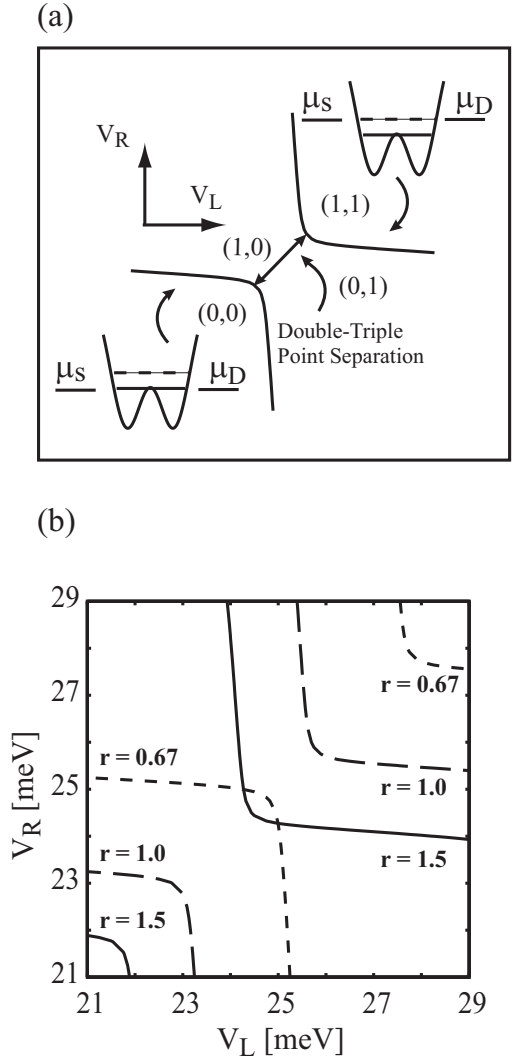}
\caption{(a) Schematic of the charge stability diagram of coupled quantum dot systems illustrating the charging process of the first and second electrons. (b) Contour plots of chemical potentials $\mu(1)$ (lower branches) and $\mu^S(2)$ (upper branches) at $r=R_y/R_x=0.67$ (dotted), $1$ (dashed) and $1.5$ (solid). $d=60$ nm, $R_x=30$ nm, $B=0$ T, reference value $\mu(1)= \mu(2)=-17$ meV.  
}
\label{fig:fig2}
\end{figure}

Figure \ref{fig:fig2}(a) shows the schematic of the charge stability diagram of coupled quantum dot systems containing up to two electrons \cite{Wiel}. $V_L$ ($V_R$) denotes the controlling gate bias for the left (right) dot. The lower curved branch (solid line) corresponds to the bias condition under which the chemical potential of the first electron in the quantum dot [$\mu(1)$, shown by the solid line inside the confining potential] is aligned with the source and drain chemical potentials $\mu_S$ and $\mu_D$ and the first electron enters the system. Here, we consider the situation close to equilibrium, so in a first approximation we assume $\mu_S=\mu_D$. When $V_L$ and $V_R$ are increased, the chemical potential for the second electron [$\mu(2)$, shown by the dotted line inside the confining potential] are lowered and aligned with $\mu_S=\mu_D$ and the second electron is charged into the coupled dots. The upper curved branch (solid line) corresponds to this situation. The separation along the main diagonal between the lower and upper branches is called the double-triple point (DTP) separation (also called the anti-crossing separation) \cite{Wiel, Hatano, Zhang}. The number in the parentheses on the left (right) gives the electron number on the left (right) dot in distinct stable charge regions.

In our simulation, we choose a fixed reference value of the source and drain chemical potentials [$\mu(1)=\mu^S(2)= -17$ meV], and plot on the same diagram the chemical potential contours for the first and second electrons with respect to two controlling gate biases so as to obtain the charge stability diagram. Figure \ref{fig:fig2}(b) shows the chemical potential contours for the first electron $\mu(1)$ (lower branches) and for two electron singlet $\mu^S(2)$ (upper branches) for $d=60$ nm and $r=R_y/R_x=0.67$ (dotted), $1$ (dashed), and $1.5$ (solid) with $R_x$ fixed at $30$ nm at zero magnetic field. As the aspect ratio decreases, the single-particle eigenenergies increase as a result of stronger confinement in each dot, hence the chemical potential contours shift from the lower left corner to the upper right corner. The DTP separation is $\Delta V_L=\Delta V_R=2.83$, $2.94$, and $3.04$ meV for $r=0.67$, $1$, and $1.5$, respectively. According to the "classical" theory in Ref. \cite{Wiel}, a larger DTP separation signifies a stronger inter-dot coupling strength. Therefore, the inter-dot coupling strength is larger for a larger aspect ratio, which is consistent with the above result of a larger exchange energy value at zero magnetic field found for $r>1$ [see Fig. \ref{fig:fig1}(a)].

\begin{figure}[t]
\begin{center}
\includegraphics[width=2.8in]{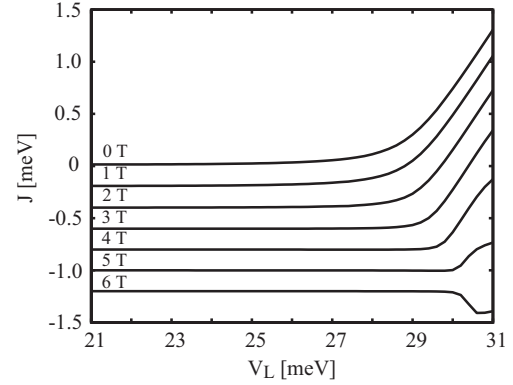}
\caption{Exchange energy $J$ as a function of $V_L$ in the model potential at different
magnetic fields. $V_R$ is fixed at $21$ meV. Other parameters are $d=60$ nm.
$R_x=R_y=30$ nm. For clarity, curves for different $B$ fields are shifted vertically
by $-0.2$ meV.}
\label{fig:fig3}
\end{center}
\end{figure}

In Fig. \ref{fig:fig3}, the exchange energy is shown as a function of 
$V_L$ at $V_R=21$ meV, or in other words, as a function of inter-dot
detuning $\epsilon=V_L-V_R$ for coupled circular dots ($R_x=R_y=30$ nm, 
$d=60$ nm) at different magnetic fields. The exchange energy at different
magnetic fields is almost invariant over the $V_L$ interval from 21 to 29 meV. 
But for  $V_L>29$ meV, it increases superlinearly for magnetic field $0\leq B\leq 5$ T 
and decreases superlinearly at $B=6$ T. 
For $B=5$ and 6 T, the $J$ saturation  is seen as well.
The onset of superlinearity is consistent with the difference in the 
electron density between the singlet and triplet states. As shown below in column (II) of 
Fig. \ref{fig:fig5}, as $V_L$ becomes larger, the electron density of the
singlet state becomes localized in the left dot while the electron
density of the triplet state continues to be spread over the two dots. The
difference in electron density and the corresponding elecron-electron interaction becomes larger with the increasing inter-dot detuning, and hence, the magnitude of the exchange energy increases.

\begin{figure}[t]
\begin{center}
\includegraphics[width=2.8in]{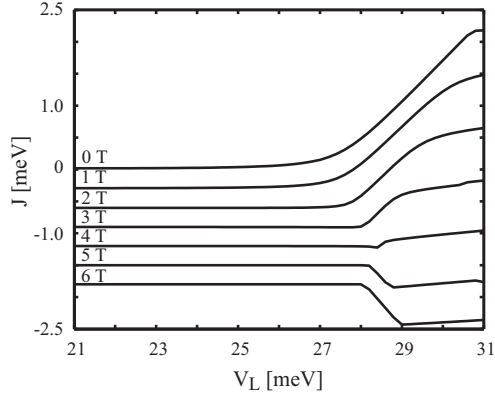}
\caption{Exchange energy $J$ as a function of $V_L$ in the model potential at different
magnetic fields. $V_R$ is fixed at $21$ meV. Other parameters are $d=60$ nm,
$R_x=30$ nm, $R_y=45$ nm. For clarity, curves for different $B$ fields are shifted vertically
by $-0.3$ meV.}
\label{fig:fig4}
\end{center}
\end{figure}

\begin{figure}
\begin{center}
\includegraphics[width=3in]{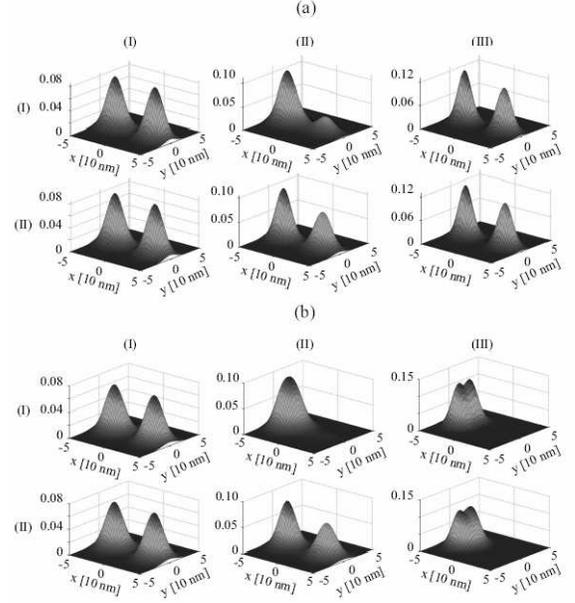}
\caption{Two-dimensional electron density plots 
for (a) $r=R_y/R_x=1$ and (b) $r=R_y/R_x=1.5$.
In each case rows (I) and (II) are for the singlet and triplet states, respectively.
Different columns are for different parameters: (I) $V_L=V_R=21$ meV, $B=0$, 
(II) $V_L=29$, $V_R=21$ meV, $B=0$ and (III) $V_L=29$, $V_R=21$ meV, $B=6$ T. For
all cases, $d=60$ nm and $R_x=30$ nm.}
\label{fig:fig5}
\end{center}
\end{figure}

Figure \ref{fig:fig4} shows the dependence of the exchange energy 
on $V_L$ at $V_R=21$ meV, for coupled elliptic dots ($R_x=30$ nm, 
$R_y=45$ nm, $d=60$ nm) at different magnetic fields. The qualitative behavior
of the exchange energy for $r=1.5$ is similiar to that for $r=1$ in 
Fig. \ref{fig:fig3}, except that saturation occurs
for all the investigated magnetic fields at smaller inter-dot detuning.
In general, the onset of exchange 
energy saturation occurs when
the inter-dot detuning is so large that the electrons in both 
singlet and triplet states starts to localize in the same dot and the
difference in the electron density between these states becomes smaller.
We observe that exchange 
energy saturation occurs for $V_L>35$ ($\epsilon=V_L-V_R>14$) meV 
for $r=1$ in the absence 
of the magnetic field (not shown).  
This detuning value for the onset of exchange energy saturation
is larger than that for the $r=1.5$ case because 
for $r=1.5$ the effective size of each dot is larger and 
the electron density for both the singlet and triplet states more easily localize into a single dot.  
The general trend in the superlinear
increase of the exchange energy as a function of the inter-dot detuning is
consistent with recent experimental observations \cite{Petta}. 
For $d=60$ nm, the exchange energy in the saturation 
regime is in the meV range, which
is typical of a single dot of the same size ($R=30$ nm)~\cite{Harju, Dmm} 
and two orders of magnitude larger than the value 
at $\epsilon=0$.

In Fig.~\ref{fig:fig5}, we show the two-dimensional electron density plots
for (a) $r=R_y/R_x=1$ and (b) $r=R_y/R_x=1.5$.
In each case rows (I) and (II) are for the singlet and triplet states, respectively.
Different columns are for different parameters: (I) $V_L=V_R=21$ meV, $B=0$, 
(II) $V_L=29$, $V_R=21$ meV, $B=0$ and (III) $V_L=29$, $V_R=21$ meV, $B=6$ T. In
all cases, $d=60$ nm and $R_x=30$ nm. By comparing the data in Fig. \ref{fig:fig5} (a) and (b), we see that at zero interdot detuning ($V_L=V_R$) and $B=0$ [column (I)], the electron density
of both the singlet and triplet states shows two peaks localized at the center of 
each dot which are symmetric with respect to $x=0$. The exchange energies for in these cases are comparable to each other ($J=15$ and $J=19$ $\mu$eV, respectively).
At non-zero interdot detuning ($V_L-V_R=8$ meV) and $B=0$ [column (II)],
the singlet electron density in (a) for $r=1$ shows a secondary peak in the right dot, while this peak is absent for a larger aspect ratio $r=1.5$ in (b). This is because for a larger aspect ratio, as was already mentioned above, the onset value of the inter-dot detuning for two electron localization into a single dot is smaller (see discussion on Figs.~\ref{fig:fig3} and \ref{fig:fig4} above) due to the relaxation of the confinement potential in the $y$ direction. In this case, the triplet electron density profiles for both cases show a secondary peak in the right dot, and are similar to one another while the calculated exchange energies are $J=297$ and $J=1050$ $\mu$eV, respectively.
At non-zero interdot detuning ($V_L-V_R=8$ meV) and $B=6$ T [column (III)],
there is a drastic difference in the electron density profiles 
between the results in (a) and (b): in (a) for both singlet
and triplet the electron density shows a higher (lower) peak localized at 
the center of the left (right) dot, while in (b), for both singlet
and triplet the electron density is totally localized in the left dot
with two equal peaks at $y=\pm8$ nm, $x=-30$ nm. Due to the large overlap of the
electron wavefunctions in the latter case, the exchange interaction $J=-577$ $\mu$eV
is more than two orders of magnitude larger than $J=-1$ $\mu$eV in the former case. 

\section{Conclusion}

We have computed the exchange energy and derived the stability diagram for model coupled elliptic quantum dot systems with different aspect ratios. We find larger exchange energy and smaller magnetic field at singlet-triple transition for dots with larger aspect ratios $R_y/R_x>1$,
which provides a better control of the exchange interaction. By investigating the double-triple point separation and the curvature of chemical potential contour lines on the stability diagram for different dot configurations, we find the inter-dot coupling decreases with increasing inter-dot separations, increasing magnetic fields, and decreasing dot aspect ratios. In the weak coupling regime (inter-dot distance $60$ nm), we find three distinctive regions in the dependence of the
exchange energy ($J$) on the inter-dot detuning ($\epsilon$): (1) For small $\epsilon$, $J$ maintains a nearly 
constant value and the electron density of the singlet and triplet states are spread between the two dots and their difference is small. (2) For intermediate $\epsilon$, $J$ depends superlinearly on $\epsilon$. Singlet electron density is well localized into one dot, while triplet electron density is still spread between the two dots. (3) For large $\epsilon$, the increase of $J$ tends to saturate and both singlet and triplet electron densities begin to localize into one dot. At a fixed aspect ratio, the onset of the superlinear and 
saturation regions depends on the magnetic field. For a constant magnetic field, the onset of these two regions
occurs at a smaller inter-dot detuning value for dots with larger aspect ratios.

\section*{Acknowledgment}

This work is supported by DARPA QUIST program through ARO Grant DAAD 19-01-1-0659. The authors thank the Material Computational Center at the University of Illinois through NSF Grant DMR 99-76550. LXZ thanks the Computer Science and Engineering Fellowship Program at the University of Illinois for support.

\end{document}